\documentclass[amsmath,amssymb,aps,prl,twocolumn,groupedaddress,superscriptaddress,10pt,reprint,longbibliography]{revtex4-2}
\usepackage{graphicx}
\usepackage{textcomp} 

\usepackage{exscale}

\usepackage{relsize}

\usepackage[colorlinks,linkcolor=blue,citecolor=blue,urlcolor=blue]{hyperref}

\begin{document}
	\title{Sorting light's radial momentum and orbital angular momentum with a parabola-like lens}
	
\author{Yuan Li}
\affiliation{Department of Physics, Xiamen University, Xiamen 361005, China}
\author{Ye Xing}
\affiliation{Department of Physics, Xiamen University, Xiamen 361005, China}
\author{Wuhong Zhang}
\email{zhangwh@xmu.edu.cn}
\affiliation{Department of Physics, Xiamen University, Xiamen 361005, China}
\affiliation{Jiujiang Research Institute, Xiamen University, Jiujiang 332199, China}
\author{Lixiang Chen}
\email{chenlx@xmu.edu.cn}
\affiliation{Department of Physics, Xiamen University, Xiamen 361005, China}

	
	\date{\today}

\begin{abstract}
The orbital angular momentum and radial momentum both describe the transverse momentum of a light field. Efficient discriminating and sorting the two kinds of momentum lies at the heart of further application. Here, we propose a parabola-like lens that can transform the orbital angular momentum and the radial momentum into different positions in the parabolas. We experimentally characterize the performance of our implementation by separating individual angular and radial momentum as well as the multiple superposition states. The reported scheme can achieve two kinds of transverse momentum identification and thus provide a possible way to complete the characterization of the full transverse momentum of an optical field. The proposed device can readily be used in multiplexing and demultiplexing of optical information, and in principle, achieve unit efficiency, and thus can be suitable for applications that involve quantum states of light.

\end{abstract}

\maketitle

The spatial modes of light give access to an in principle unbounded state space and have been established as a resource for encoding information in higher-dimensional alphabets beyond one bit per photon. Mode-divided multiplexing (MDM) is such a technique and has become the next frontier in optical communication \cite{Richardson2013SpaceDivision,li2014space,trichili2019communicating}. The orbital angular momentum (OAM), as one part of the transverse momentum of a light field, has been widely studied for optical communication \cite{wang2012terabit,bozinovic2013terabit,willner2015optical}. The radial momentum (RM), as the other part of the transverse momenta, is less well known and has recently been studied. Using the radial position and RM variables, the Einstein–Podolsky–Rosen paradox relation can be observed on this new platform \cite{chen2019realization}, a radial version of light diffraction can be well described by the newly introduced RM variables \cite{ma2020radial}, even the uncertainty principle can also be invested based on these two radial degrees of freedom \cite{zhang2022experimental}. These works indicate that RM has played a role in the fundamental test of quantum information and is possible to be used for optical communication. One of the fundamental and priority problems is how to discriminate and sort the OAM and RM when the light field contains both momenta.

It is noted that the OAM sorting techniques have been well studied with the interferometric method \cite{leach2002measuring,leach2004interferometric,zhang2014mimicking} and diffraction technique \cite{berkhout2008method,hickmann2010unveiling} in a few decades ago. Some interesting methods, such as the Cartesian to log-polar coordinate transformation \cite{berkhout2010efficient, mirhosseini2013efficient}, spiral transformation \cite{wen2018spiral}, deep-learning method \cite{liu2019superhigh} are the other choices for OAM sorting. However, little attention has been paid to sort the RM. Some radial mode sorters are proposed to sort the two indexes of the Bessel-Gauss (BG) modes or Laguerre-Gaussian (LG) modes. Early in 2013, Dudley \textit{et al.} proposed a method to efficiently sort the different transverse wave number $k_r$ and azimuthal mode index $\ell$ of the BG modes \cite{dudley2013efficient}. Then a digital axicons was proposed to detect the arbitrarily radial and azimuthal indices of the BG beams \cite{trichili2014detection}.
The possibility of sorting the radial index of the LG modes was proposed using a random scatter \cite{fickler2017custom} and then optimized with two phase transformation screens \cite{fickler2020full}. Then the azimuthal modes and radial modes of the LG beam could be efficiently sorted at the same time by considering the accumulated Gauy phases in the LG mode \cite{gu2018gouy} and using the fractional Fourier transform lenses \cite{zhou2017sorting} and even using the polarization-dependent fractional Fourier transform lens \cite{fu2018realization}. Furthermore, the LG mode sorter based on multiplane light conversion was proposed to reach 210 mode sorting \cite{fontaine2019laguerre}. 
It should be noted that the RM is different from the radial mode of an LG or BG beam. Since the radial mode does not refer to the traditional momentum \cite{plick2015physical}, one can distinguish the radial modes from the RM using the RM operator: $\hat{p_r}=-i\hbar(\frac{\partial}{\partial\rho}+\frac{1}{2\rho})$ \cite{chen2019realization}. The average RM of a light field can be calculated by this operator and should be equal to the RM $p_r$. If a light field contains the radial modes but $p_r=0$, one can define that the light field carries zero RM. Furthermore, since the beam that contains RM ($p_r\neq0$) usually undergoes significant changes in spot size during propagation, the above radial mode sorter, which needs incident beam propagation between different elements, is not applicable to RM sorting, see the Supplementary Material (SM) for detailed demonstration. 

Here, we propose a parabola-like lens that can transform OAM and RM into different positions. It can be thought of as the case for the log-polar azimuthal-only mode sorter compared to the other coordinate transformation mode sorters \cite{berkhout2010efficient,mirhosseini2013efficient,wen2018spiral}. Similar log-polar azimuthal-only mode sorters have been proposed as the angular lens to separate different OAMs into radial positions by Sahu \textit{et al.} in 2018 \cite{sahu2018angular} and have been optimized to perform a better OAM-separation effect \cite{zhou2022orbital,lv2022sorting}. However, the existing phase distribution of the angular lens also has the defect that it cannot separate OAM and RM at the same time. By considering the ring shape of the different RM beams, we propose a new phase distribution which can modulate the input OAM and RM into the scatter of points in different parabolas. Our theoretical and experimental results, which demonstrate the ability to sort individual OAM and RM states that reside in the fifteen-dimensional state space successfully, illustrate a good performance of the parabola-like lens. Our proposal holds promise for multiplexing and de-multiplexing optical information to further boosting the information capacity of a single photon.

To understand the specifics of our implementation, we start by describing the light field, which contains both RM and OAM. 
\begin{figure}
\includegraphics[width=1.0\columnwidth]{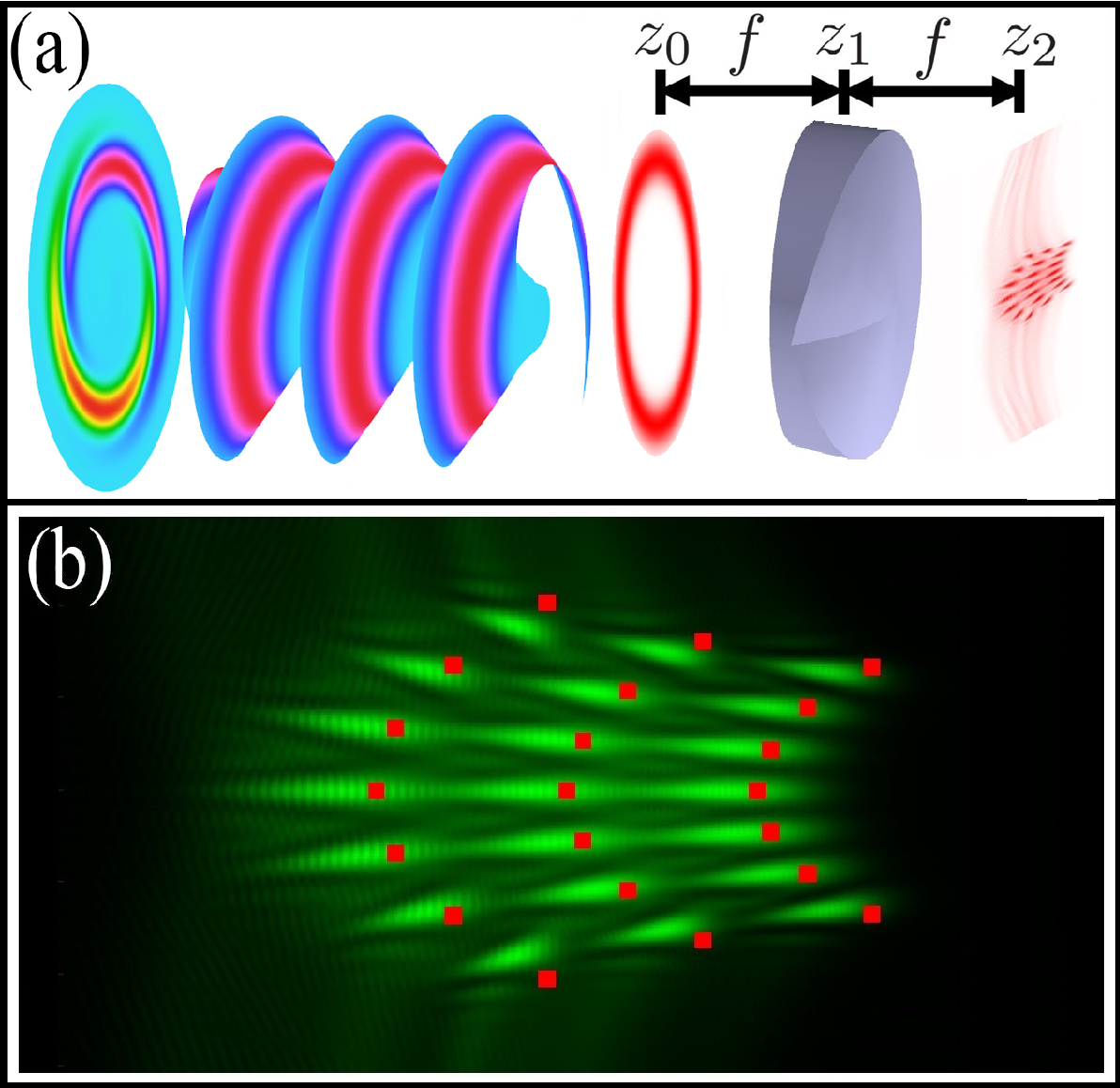}
\caption{(a) Diagrammatic sketch of the parabola-like lens for sorting the RM and OAM. (b) Green light patterns are the theoretical simulation of sorting 21 modes with the above lens, while the red spots denote the analytical mapping position $(x,y)$ in $z_2$ plane for the different input OAM and RM ($\ell,p_r$) modes at $z_0$ plane with $C=0.03$.}
\label{fig1}
\end{figure}
In cylindrical coordinate system, the OAM defines the photon's azimutal ($\phi$) momentum distribution and have been well studied, while the photon's RM describes that the transverse momentum can be also distributed along the $\rho$ direction. Such the beam will expand during propagation with RM $p_r>0$ otherwise it will shrink with $p_r<0$. Our recent demonstrations have shown that one can use the conical phase to produce a light field with RM \cite{chen2019realization,ma2020radial,zhang2022experimental}. So we can define a light field that contain both of the OAM and RM beam at $z=z_0$ plane as:
\begin{equation}
\begin{split}
E_{\ell,p_r}(\rho,\phi)=(\frac{\sqrt{d}\rho}{R_0})^{d}\exp(-\frac{d\rho^2}{2{R_0}^2})\exp(i\ell\phi)\exp(\frac{i{p_r}\rho}{\hbar}),
\label{eq1}
\end{split}
\end{equation}
where $(\frac{\sqrt{d}\rho}{R_0})^{d}\exp(-\frac{d\rho^2}{2{R_0}^2})$ describes a specified designed amplitude modulation to make the beam  with annular intensity, $R_0$ is the radius of the annular beam, $d$ decides the width of the ring, the higher $d$ the narrower of the ring with the same radius $R_0$, $\ell$ is the OAM quantum number and $p_r$ is the RM of each photons in the light field. As shown in Fig. \ref{fig1}(a), we plot the electric field, cophasal surface, and intensity distribution of the light field with $\ell=1$, $p_r=30000\hbar m^{-1}$ at one moment. Since RM is a continuous variable, the value of RM is set as the average value of the modes generated by Eq. (1). From Eq. (1), the RM distribution of the beam can be calculated, which has a very narrow peak and the FWHM of the peak is less than 1\% of the average value.
Compared with the spiral cophasal surface of the OAM beam, the beam produced here with both OAM and RM $p_r>0$ has a special spiral cone-shaped cophasal surface. Due to the existing positive RM and OAM, the transverse intensity distribution of the beam exhibits a ring-shaped structure and the beam will have a cone-shaped intensity expanding during propagation. It is noted that with different $p_r$, the information carrier channel will be boosted exponentially compared to only considering the OAM beam. So, how to efficiently sort the OAM and RM lies at the heart of further application. By considering the expanding effect of the beam that contains both RM and OAM from the initial plane to the observe plane, we propose a pure phase element that is positioned at the center ($z_1$) between the input plane ($z_0$) and the output plane ($z_2$) to perform the two-kind momentum transformation, as illustrated in Fig. \ref{fig1}(a). The pure phase element can act as a lens, the focal length of which is $f$ and its phase distribution can be deduced as:
\begin{equation}
\begin{split}
T(\rho,\phi)=\exp[-i\frac{k\rho(2\rho-R_0)}{2f}\phi^{2}],
\label{eq2}
\end{split}
\end{equation}
where $R_0$ denote the beam radius at input plane ($z_0$), $R_1$ is the beam radius at the lens plane ($z_1$). As long as the radius of the distribution of the annular light intensity in the input plane is $R_0$, the beam with different OAM and RM can be well separated and focused in the observe plane ($z_2$). While if the input beam has propagation with the new beam radius $R_0'$, the phase distribution can be recalculated based on the Eq. (\ref{eq2}) to realize the separation, see more in the SM.
To simulate the propagation behavior of the beam after the lens, we consider the propagation equation as follows:
\begin{equation}
\begin{split}
E(\rho',\phi',z')=&\frac{\exp(ikz')}{i\lambda z'}e^{\frac{ik{\rho'}^2}{2z'}}\int_{0}^{+\infty}\int_{0}^{2\pi}E(\rho,\phi,z_1)\\&T(\rho,\phi)e^{\frac{ik\rho^2}{2z'}}e^{\frac{-ik\rho\rho' cos(\phi'-\phi)}{z'}}\rho d\rho d\phi,
\label{eq3}
\end{split}
\end{equation}
where $E(\rho,\phi,z_1)$ is the input light field after propagation at a distance $f$ and can be integrated based on Eq. (\ref{eq3}) with $T(\rho,\phi)=1$. The simulated separation results of 21 different modes with combined 3 RM values and 7 OAM values in the observe plane are shown in Fig. \ref{fig1}(b). The input beam with different $\ell$ and the same $p_r$ distributed on a parabola, and different RM $p_r$ distributed on different parabola, indicating that our proposed lens can focus these modes very well and is concentrated with no overlap.
\begin{figure}
\includegraphics[width=1.0\columnwidth]{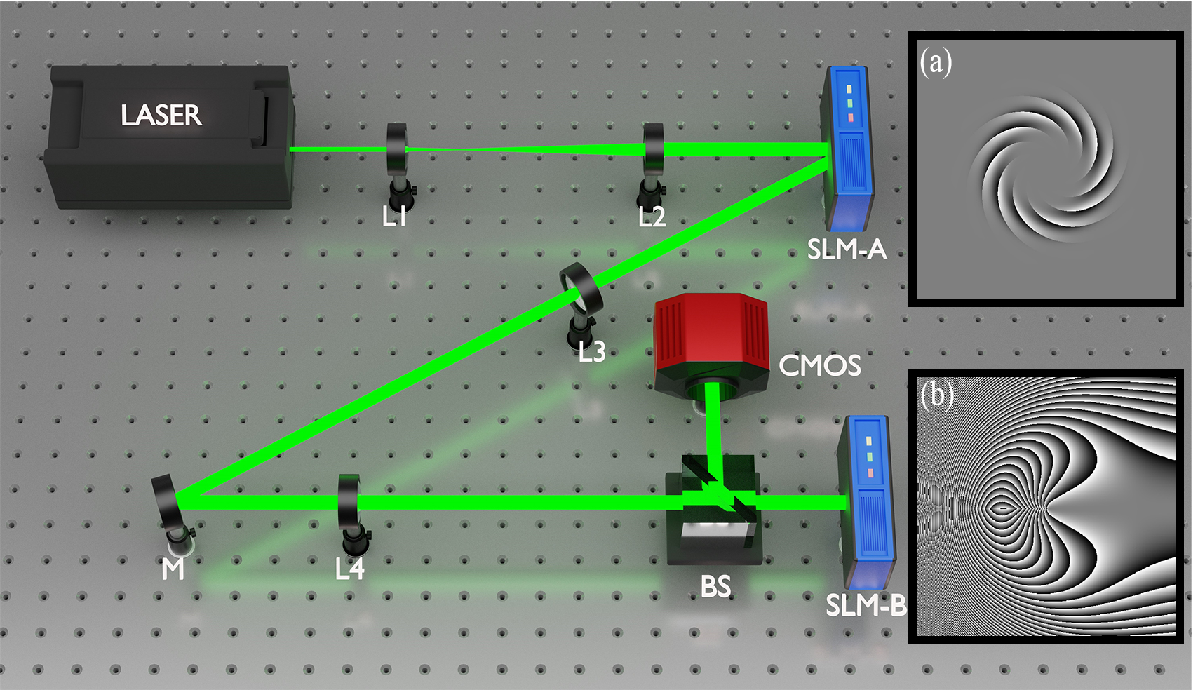}
\caption{Experiment setup. L:lens, M: mirror, BS: beam splitter. (a) the desired phase mask used to produce RM and OAM. (b) the phase distribution of the parabola-like lens.}
\label{fig2}
\end{figure}

To clearly show the parabola property of the lens, we also analyze the separation principle with the photonic momentum transformation (PMT) technique \cite{guo2021spin}. When an incident beam contain both the RM $p_r$ and OAM $\ell\hbar$, the focused positions of different momentum on observation plane in Cartesian coordinate can be written as:
\begin{equation}
    \begin{split}
        &x=R_{0}+\frac{2f p_r}{k\hbar}+\frac{\ell^2 f}{2k(R_0+\frac{f p_r}{k\hbar})}C
        \\&y=\frac{\ell f}{k(R_0+\frac{f p_r}{k\hbar})},
        \label{eq4}
    \end{split}
\end{equation}
where $C$ is a constant, which comes from mathematical analyzing with the PMT, the PMT method did not limit the value of C but we can get a value from comparison with numerical simulation and experimental result. Based on Eq. (\ref{eq4}), we can calculate the coordinate $(x,y)$ of the different modes ($\ell,p_r$). We mark the analytical focus position of each mode with a red point shown in Fig. \ref{fig1}(b). The spot with the same $p_r$ and different $\ell$ is arranged on a parabola, while with increasing $p_r$, the position goes right because the quadratic coefficient of the parabola increases. In addition, the distance between different $\ell$ is closer. The red points calculated by Eq. (\ref{eq4}) are mostly well placed in the centered of the patterns simulated by Eq. (\ref{eq3}), which exactly prove the parabola property of our proposed parabola-like lens.

\begin{figure}[t]
\includegraphics[width=1\columnwidth]{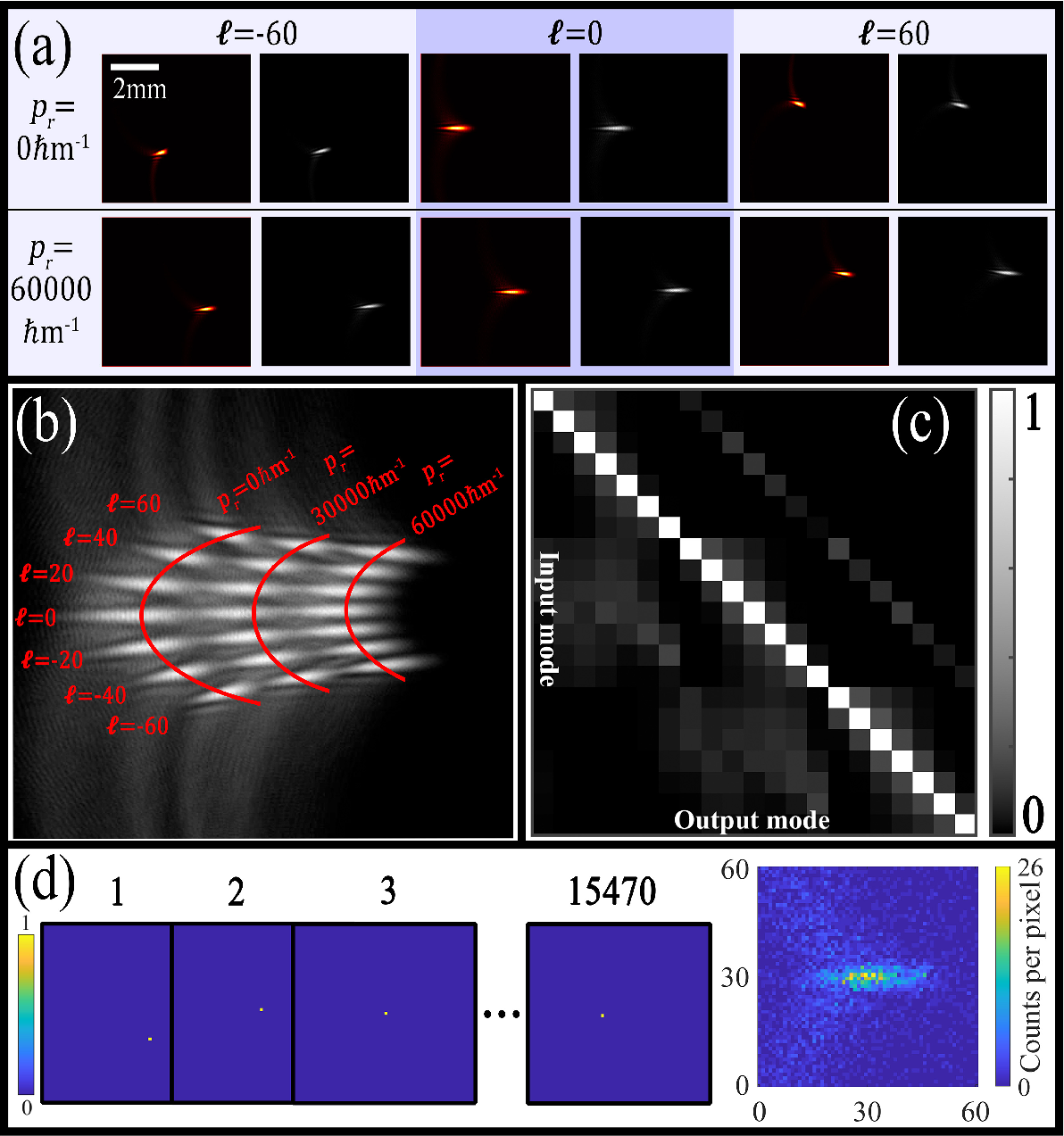}
\caption{Single OAM and RM mode sorting with the parabola-like lens. (a) Red spot denotes the simulation results, while the gray spots represent the experimental captured intensity. (b) Superposition of all the gray spots in one image. (c) is the crosstalk measured matrix for the different input modes.(d) Single photon images captured by ICCD camera and the photon distribution obtained by accumulating 15470 single photon images for the input mode $\ell=0,p_r=0$.}
\label{fig3}
\end{figure}

To test our proposed lens, we build an easy-to-implement optical system as schematic overview in Fig. \ref{fig2}. A 532nm laser is expanded by L1=50mm, L2=200mm and incident to spatial light modulator A (SLM-A). The SLM-A loads the desired hologram grating to produce the input light field as shown in Eq. (\ref{eq1}). The phase mask of the hologram is shown in Fig. \ref{fig2} (a). Then, the desired beam which contains both the OAM and the RM is obtained in the image plane of the 4F filtering system (L3=200mm, L4=200mm). The distance between the image plane and SLM-B is just $f$ that calculated by Eq. (\ref{eq2}). We used a beam splitter (BS) to ensure normal incidence in SLM-B. The SLM-B is used to produce the desired parabola-like lens of which the phase distribution is shown in Fig. \ref{fig2}(b).
The reflected light is divided by BS and received by a CMOS camera positioned at a distance of $f$ from SLM-B. To test the lens can be also worked on in photon count level, we add a series of neutral density filters to reduce the input light in photon count level and use an intensified charge-coupled device (ICCD) camera to capture the photon distribution (see more in SM).

\begin{figure}[t]
\includegraphics[width=\columnwidth]{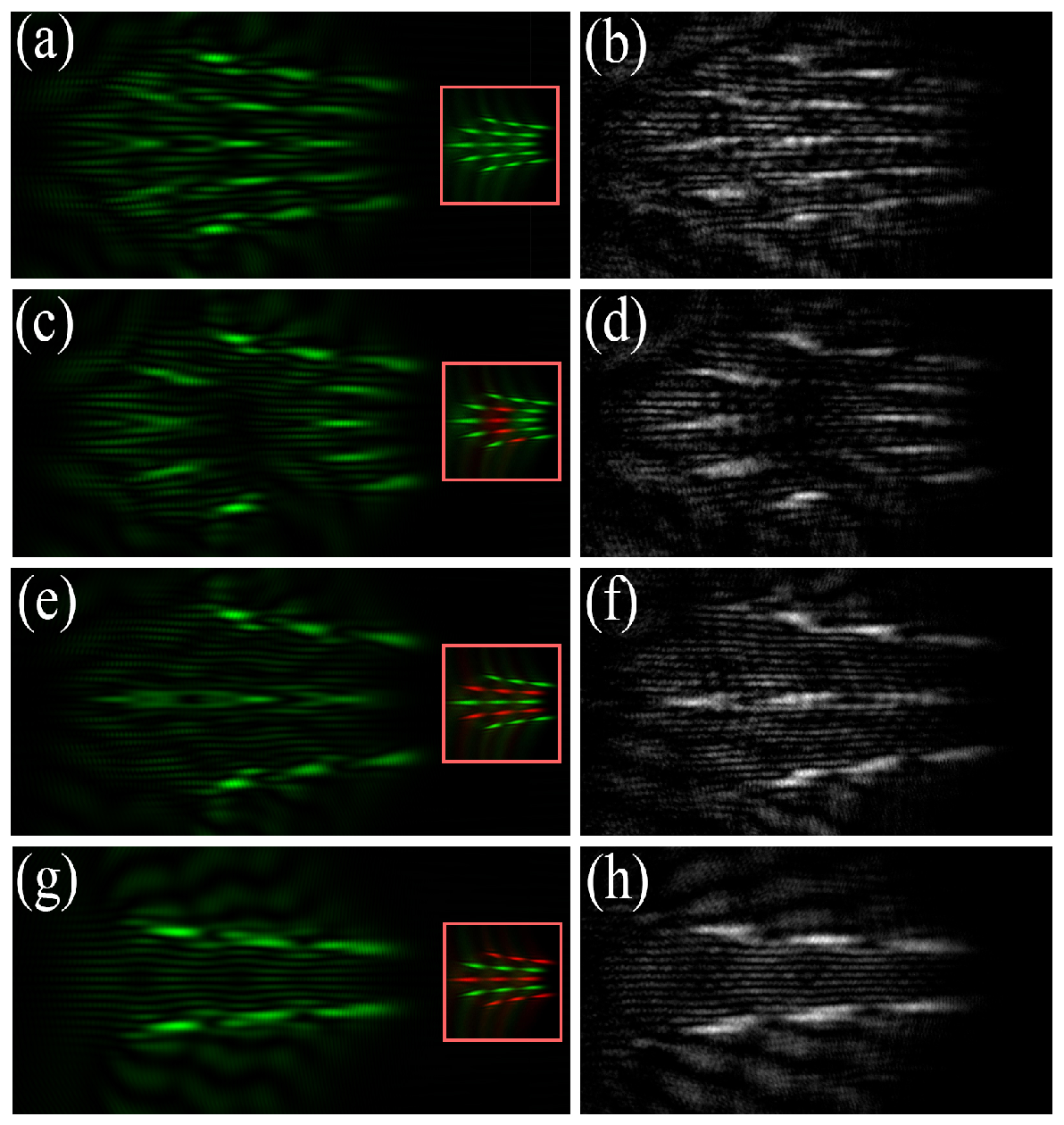}
\caption{Multiple OAM and RM modes superposition sorting effect with the parabola-like lens.}
\label{fig4}
\end{figure}
Firstly, we test the single mode separation performance of the lens. The simulation and experimental results are shown in Fig.\ref{fig3}(a). We show 6 different modes, which combine 2 RM values ($p_r=0,60000\hbar m^{-1}$) and 3 OAM values ($\ell=0,\pm60$) as input light fields. One can see that each of the input modes is focused in different positions as predicted by Eqs. (\ref{eq3}) and (\ref{eq4}). Then we superposition of all the captured single mode into one image, the same RM but different OAM are distributed on the same parabola but different RM are distributed in different parabola, as the red parabola line clearly shows in Fig. \ref{fig3} (b). To further quantitatively investigate the performance of the proposed parabola-like lens, we measure the intensity of the mode and plot the crosstalk matrix for the different input modes, as shown in Fig. \ref{fig3}(c). The crosstalk is measured by putting an intensity mask in the observer plane and recording the power of all 21 modes. Each mode has its non-overlapping transparency light area, so the ratios of each mode are defined as the total power transmitted from the mask of the corresponding modes. One may further optimize the cross-talk by utilizing the usual material instead of the SLM for making the parabola-like lens. 
To test the ability of our lens to work at the single-photon level, the single-photon images captured by ICCD are illustrated in Fig.\ref{fig3} (d). By summing 15,470 single-photon images, we can obtain the predicted modes at the right position. So, by setting the pixel area of the different input modes, our lens can be used to sort the OAM and RM in single-photon level(see more analyses in SM ).
\begin{figure}
\includegraphics[width=1.0\columnwidth]{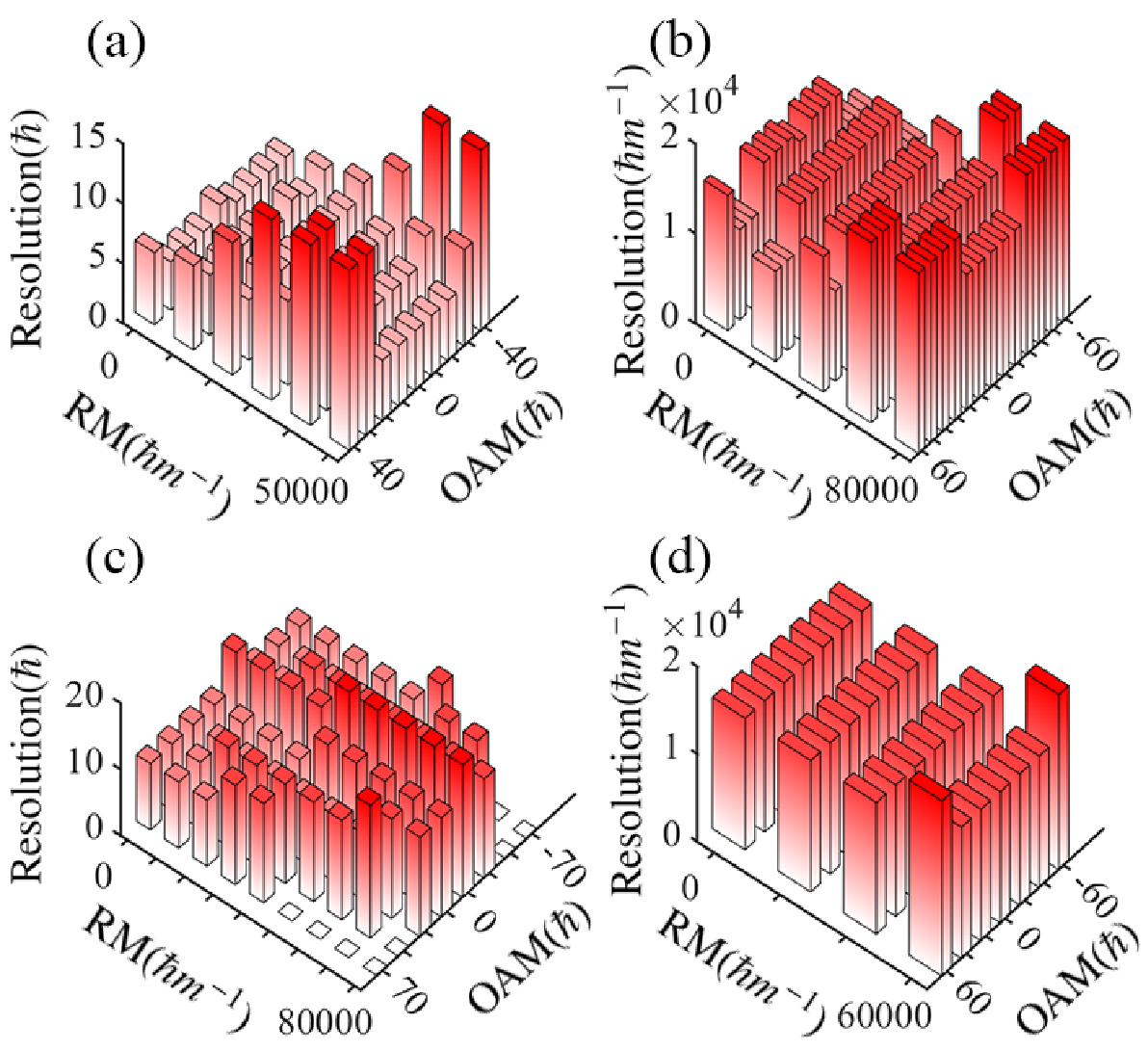}
\caption{Experimental test of the resolution distribution for OAM  and RM with the parabola-like lens. (a) and (b) denote the OAM and RM resolution distribution for single mode input while (c) and (d) denote the superposition modes input. Zero data means no test.}
\label{fig5}
\end{figure}

To test the separation performance of multiple superposition modes with the lens, we superpose 15 modes that combine 3 RM values ($p_r=0,30000\hbar m^{-1},60000\hbar m^{-1}$) and 5 OAM values ($\ell=0,\pm30,\pm60$) as input modes. The intensity captured by the CMOS camera is shown in Fig. \ref{fig4}(b), while the numerical simulation results are illustrated in Fig. \ref{fig4}(a). The inset in the upper right corner shows the filtered $15$ modes. Due to the diffractive optical property of the desired lens, we observed some scattered light around each of the separation modes. Although the shape of each spot is affected by the adjacent modes, each mode is positioned at the right place as desired. The scatter effect can be further suppressed by reducing the number of input modes. We reduce the input modes to $11, 9, 6$, as the green spot shown in Figs. \ref{fig4}(c),(e),(g), while the red spots of the inset denote the modes we reduced. The separation effect is getting better and better as the number of input modes is reduced, as shown in Figs. \ref{fig4}(d),(f),(h). We also test the performance of our lens for sorting the superposition modes at the single-photon level in SM. Our experimental results, which demonstrate the ability to sort superposition OAM and RM states successfully, illustrate the good performance of the proposed parabola-like lens. By loading information into the different input modes with a single input beam, our proposed device can readily be used as a demultiplexer of optical information, and in principle with unit efficiency. 

It is noted that resolution is a key performance for an optical information demultiplexer. To study the resolution of our parabola-like lens, we further conduct a series of experimental tests divided into single-mode input and superposition mode input. For single mode input, as illustrated in Fig. \ref{fig5} (a), the OAM resolution distribution is around $4\hbar-15\hbar$, while in Fig. \ref{fig5} (b), the RM resolution distribution is around $10000 \hbar m^{-1}-20000\hbar m^{-1}$. For input of superposition modes, the OAM resolution distribution is around $10\hbar-20\hbar$ and the RM resolution distribution is around $15000 \hbar m^{-1}-20000\hbar m^{-1}$, as shown in Figs. \ref{fig5} (c) and (d). The resolution for single mode is significantly better than multimode superposition input due to the interference effect between the adjacent modes on the observation plane. Besides, based on Eq. (\ref{eq4}), the focus position is proportional to the square of OAM, the resolution of OAM decreases as the RM increases. However, we find that the smaller $R_0$ and the larger $f$ of the desired lens will improve the OAM resolution, while the larger $f$ will improve the RM resolution. The experimental test, resolution calculation, and influence of parabola-like lens setting parameters on the resolution are detailed in the SM.

In conclusion, we have designed a parabola-shaped lens and carefully analyzed the relation between the coordinate $(x,y)$ in the Fourier plane with the different input OAM and RM ($\ell,p_r$) modes in the input plane. The special lens has shown an outstanding performance for sorting the OAM and RM residing in a light field or even at the single-photon level, which holds promise for multiplexing and demultiplexing optical information to further boosting the information capacity of a single photon. Based on this concise single element, one can also achieve the two kinds of transverse momentum identification, and thus provide a possible way to complete characterization of the full transverse momentum of an optical field. It is also noted that the achievements in the spatial mode multiplexing technique with meta-surface have been intensively studied in recent years \cite{wangmetasurface, ahmed2022optical,cheng2022ultracompact, guo2021spin}, our proposed parabola-like lens may also be manufactured with a single meta-surface to implement further optical integration.

\begin{acknowledgments}
This work is supported by the National Natural Science Foundation of China (12374280, 12034016, 12192254, 92250304), the National Key R\&D Program of China (2023YFA1407200, 2023YFA1407203), the Natural Science Foundation of Fujian Province (2021J02002, 2023J01007), the Fundamental Research Funds for the Central Universities at Xiamen University (20720220030), Xiaomi Young Talents Program, Jiujiang Xun Cheng talents Program, and the program for New Century Excellent Talents in University of China (NCET-13-0495).
 \end{acknowledgments}


\begin{thebibliography}{34}%
\makeatletter
\providecommand \@ifxundefined [1]{%
 \@ifx{#1\undefined}
}%
\providecommand \@ifnum [1]{%
 \ifnum #1\expandafter \@firstoftwo
 \else \expandafter \@secondoftwo
 \fi
}%
\providecommand \@ifx [1]{%
 \ifx #1\expandafter \@firstoftwo
 \else \expandafter \@secondoftwo
 \fi
}%
\providecommand \natexlab [1]{#1}%
\providecommand \enquote  [1]{``#1''}%
\providecommand \bibnamefont  [1]{#1}%
\providecommand \bibfnamefont [1]{#1}%
\providecommand \citenamefont [1]{#1}%
\providecommand \href@noop [0]{\@secondoftwo}%
\providecommand \href [0]{\begingroup \@sanitize@url \@href}%
\providecommand \@href[1]{\@@startlink{#1}\@@href}%
\providecommand \@@href[1]{\endgroup#1\@@endlink}%
\providecommand \@sanitize@url [0]{\catcode `\\12\catcode `\$12\catcode `\&12\catcode `\#12\catcode `\^12\catcode `\_12\catcode `\%12\relax}%
\providecommand \@@startlink[1]{}%
\providecommand \@@endlink[0]{}%
\providecommand \url  [0]{\begingroup\@sanitize@url \@url }%
\providecommand \@url [1]{\endgroup\@href {#1}{\urlprefix }}%
\providecommand \urlprefix  [0]{URL }%
\providecommand \Eprint [0]{\href }%
\providecommand \doibase [0]{http://dx.doi.org/}%
\providecommand \selectlanguage [0]{\@gobble}%
\providecommand \bibinfo  [0]{\@secondoftwo}%
\providecommand \bibfield  [0]{\@secondoftwo}%
\providecommand \translation [1]{[#1]}%
\providecommand \BibitemOpen [0]{}%
\providecommand \bibitemStop [0]{}%
\providecommand \bibitemNoStop [0]{.\EOS\space}%
\providecommand \EOS [0]{\spacefactor3000\relax}%
\providecommand \BibitemShut  [1]{\csname bibitem#1\endcsname}%
\let\auto@bib@innerbib\@empty
\bibitem [{\citenamefont {Richardson}\ \emph {et~al.}(2013)\citenamefont {Richardson}, \citenamefont {Fini},\ and\ \citenamefont {Nelson}}]{Richardson2013SpaceDivision}%
  \BibitemOpen
  \bibfield  {author} {\bibinfo {author} {\bibfnamefont {D.}~\bibnamefont {Richardson}}, \bibinfo {author} {\bibfnamefont {J.}~\bibnamefont {Fini}}, \ and\ \bibinfo {author} {\bibfnamefont {L.}~\bibnamefont {Nelson}},\ }\href@noop {} {\bibfield  {journal} {\bibinfo  {journal} {Nat. Photonics}\ }\textbf {\bibinfo {volume} {7}},\ \bibinfo {pages} {354} (\bibinfo {year} {2013})}\BibitemShut {NoStop}%
\bibitem [{\citenamefont {Li}\ \emph {et~al.}(2014)\citenamefont {Li}, \citenamefont {Bai}, \citenamefont {Zhao},\ and\ \citenamefont {Xia}}]{li2014space}%
  \BibitemOpen
  \bibfield  {author} {\bibinfo {author} {\bibfnamefont {G.}~\bibnamefont {Li}}, \bibinfo {author} {\bibfnamefont {N.}~\bibnamefont {Bai}}, \bibinfo {author} {\bibfnamefont {N.}~\bibnamefont {Zhao}}, \ and\ \bibinfo {author} {\bibfnamefont {C.}~\bibnamefont {Xia}},\ }\href@noop {} {\bibfield  {journal} {\bibinfo  {journal} {Adv. Opt. Photonics}\ }\textbf {\bibinfo {volume} {6}},\ \bibinfo {pages} {413} (\bibinfo {year} {2014})}\BibitemShut {NoStop}%
\bibitem [{\citenamefont {Trichili}\ \emph {et~al.}(2019)\citenamefont {Trichili}, \citenamefont {Park}, \citenamefont {Zghal}, \citenamefont {Ooi},\ and\ \citenamefont {Alouini}}]{trichili2019communicating}%
  \BibitemOpen
  \bibfield  {author} {\bibinfo {author} {\bibfnamefont {A.}~\bibnamefont {Trichili}}, \bibinfo {author} {\bibfnamefont {K.-H.}\ \bibnamefont {Park}}, \bibinfo {author} {\bibfnamefont {M.}~\bibnamefont {Zghal}}, \bibinfo {author} {\bibfnamefont {B.~S.}\ \bibnamefont {Ooi}}, \ and\ \bibinfo {author} {\bibfnamefont {M.-S.}\ \bibnamefont {Alouini}},\ }\href@noop {} {\bibfield  {journal} {\bibinfo  {journal} {IEEE COMMUN SURV TUT}\ }\textbf {\bibinfo {volume} {21}},\ \bibinfo {pages} {3175} (\bibinfo {year} {2019})}\BibitemShut {NoStop}%
\bibitem [{\citenamefont {Wang}\ \emph {et~al.}(2012)\citenamefont {Wang}, \citenamefont {Yang}, \citenamefont {Fazal}, \citenamefont {Ahmed}, \citenamefont {Yan}, \citenamefont {Huang}, \citenamefont {Ren}, \citenamefont {Yue}, \citenamefont {Dolinar}, \citenamefont {Tur} \emph {et~al.}}]{wang2012terabit}%
  \BibitemOpen
  \bibfield  {author} {\bibinfo {author} {\bibfnamefont {J.}~\bibnamefont {Wang}}, \bibinfo {author} {\bibfnamefont {J.-Y.}\ \bibnamefont {Yang}}, \bibinfo {author} {\bibfnamefont {I.~M.}\ \bibnamefont {Fazal}}, \bibinfo {author} {\bibfnamefont {N.}~\bibnamefont {Ahmed}}, \bibinfo {author} {\bibfnamefont {Y.}~\bibnamefont {Yan}}, \bibinfo {author} {\bibfnamefont {H.}~\bibnamefont {Huang}}, \bibinfo {author} {\bibfnamefont {Y.}~\bibnamefont {Ren}}, \bibinfo {author} {\bibfnamefont {Y.}~\bibnamefont {Yue}}, \bibinfo {author} {\bibfnamefont {S.}~\bibnamefont {Dolinar}}, \bibinfo {author} {\bibfnamefont {M.}~\bibnamefont {Tur}},  \emph {et~al.},\ }\href@noop {} {\bibfield  {journal} {\bibinfo  {journal} {Nat. Photonics}\ }\textbf {\bibinfo {volume} {6}},\ \bibinfo {pages} {488} (\bibinfo {year} {2012})}\BibitemShut {NoStop}%
\bibitem [{\citenamefont {Bozinovic}\ \emph {et~al.}(2013)\citenamefont {Bozinovic}, \citenamefont {Yue}, \citenamefont {Ren}, \citenamefont {Tur}, \citenamefont {Kristensen}, \citenamefont {Huang}, \citenamefont {Willner},\ and\ \citenamefont {Ramachandran}}]{bozinovic2013terabit}%
  \BibitemOpen
  \bibfield  {author} {\bibinfo {author} {\bibfnamefont {N.}~\bibnamefont {Bozinovic}}, \bibinfo {author} {\bibfnamefont {Y.}~\bibnamefont {Yue}}, \bibinfo {author} {\bibfnamefont {Y.}~\bibnamefont {Ren}}, \bibinfo {author} {\bibfnamefont {M.}~\bibnamefont {Tur}}, \bibinfo {author} {\bibfnamefont {P.}~\bibnamefont {Kristensen}}, \bibinfo {author} {\bibfnamefont {H.}~\bibnamefont {Huang}}, \bibinfo {author} {\bibfnamefont {A.~E.}\ \bibnamefont {Willner}}, \ and\ \bibinfo {author} {\bibfnamefont {S.}~\bibnamefont {Ramachandran}},\ }\href@noop {} {\bibfield  {journal} {\bibinfo  {journal} {Science}\ }\textbf {\bibinfo {volume} {340}},\ \bibinfo {pages} {1545} (\bibinfo {year} {2013})}\BibitemShut {NoStop}%
\bibitem [{\citenamefont {Willner}\ \emph {et~al.}(2015)\citenamefont {Willner}, \citenamefont {Huang}, \citenamefont {Yan}, \citenamefont {Ren}, \citenamefont {Ahmed}, \citenamefont {Xie}, \citenamefont {Bao}, \citenamefont {Li}, \citenamefont {Cao}, \citenamefont {Zhao} \emph {et~al.}}]{willner2015optical}%
  \BibitemOpen
  \bibfield  {author} {\bibinfo {author} {\bibfnamefont {A.~E.}\ \bibnamefont {Willner}}, \bibinfo {author} {\bibfnamefont {H.}~\bibnamefont {Huang}}, \bibinfo {author} {\bibfnamefont {Y.}~\bibnamefont {Yan}}, \bibinfo {author} {\bibfnamefont {Y.}~\bibnamefont {Ren}}, \bibinfo {author} {\bibfnamefont {N.}~\bibnamefont {Ahmed}}, \bibinfo {author} {\bibfnamefont {G.}~\bibnamefont {Xie}}, \bibinfo {author} {\bibfnamefont {C.}~\bibnamefont {Bao}}, \bibinfo {author} {\bibfnamefont {L.}~\bibnamefont {Li}}, \bibinfo {author} {\bibfnamefont {Y.}~\bibnamefont {Cao}}, \bibinfo {author} {\bibfnamefont {Z.}~\bibnamefont {Zhao}},  \emph {et~al.},\ }\href@noop {} {\bibfield  {journal} {\bibinfo  {journal} {Adv. Opt. Photonics}\ }\textbf {\bibinfo {volume} {7}},\ \bibinfo {pages} {66} (\bibinfo {year} {2015})}\BibitemShut {NoStop}%
\bibitem [{\citenamefont {Chen}\ \emph {et~al.}(2019)\citenamefont {Chen}, \citenamefont {Ma}, \citenamefont {Qiu}, \citenamefont {Zhang}, \citenamefont {Zhang},\ and\ \citenamefont {Boyd}}]{chen2019realization}%
  \BibitemOpen
  \bibfield  {author} {\bibinfo {author} {\bibfnamefont {L.}~\bibnamefont {Chen}}, \bibinfo {author} {\bibfnamefont {T.}~\bibnamefont {Ma}}, \bibinfo {author} {\bibfnamefont {X.}~\bibnamefont {Qiu}}, \bibinfo {author} {\bibfnamefont {D.}~\bibnamefont {Zhang}}, \bibinfo {author} {\bibfnamefont {W.}~\bibnamefont {Zhang}}, \ and\ \bibinfo {author} {\bibfnamefont {R.~W.}\ \bibnamefont {Boyd}},\ }\href@noop {} {\bibfield  {journal} {\bibinfo  {journal} {Phys. Rev. Lett.}\ }\textbf {\bibinfo {volume} {123}},\ \bibinfo {pages} {060403} (\bibinfo {year} {2019})}\BibitemShut {NoStop}%
\bibitem [{\citenamefont {Ma}\ \emph {et~al.}(2020)\citenamefont {Ma}, \citenamefont {Zhang}, \citenamefont {Qiu}, \citenamefont {Chen},\ and\ \citenamefont {Chen}}]{ma2020radial}%
  \BibitemOpen
  \bibfield  {author} {\bibinfo {author} {\bibfnamefont {T.}~\bibnamefont {Ma}}, \bibinfo {author} {\bibfnamefont {D.}~\bibnamefont {Zhang}}, \bibinfo {author} {\bibfnamefont {X.}~\bibnamefont {Qiu}}, \bibinfo {author} {\bibfnamefont {Y.}~\bibnamefont {Chen}}, \ and\ \bibinfo {author} {\bibfnamefont {L.}~\bibnamefont {Chen}},\ }\href@noop {} {\bibfield  {journal} {\bibinfo  {journal} {Opt. Lett.}\ }\textbf {\bibinfo {volume} {45}},\ \bibinfo {pages} {5152} (\bibinfo {year} {2020})}\BibitemShut {NoStop}%
\bibitem [{\citenamefont {Zhang}\ \emph {et~al.}(2022)\citenamefont {Zhang}, \citenamefont {Zhang}, \citenamefont {Qiu}, \citenamefont {Chen}, \citenamefont {Franke-Arnold},\ and\ \citenamefont {Chen}}]{zhang2022experimental}%
  \BibitemOpen
  \bibfield  {author} {\bibinfo {author} {\bibfnamefont {Z.}~\bibnamefont {Zhang}}, \bibinfo {author} {\bibfnamefont {D.}~\bibnamefont {Zhang}}, \bibinfo {author} {\bibfnamefont {X.}~\bibnamefont {Qiu}}, \bibinfo {author} {\bibfnamefont {Y.}~\bibnamefont {Chen}}, \bibinfo {author} {\bibfnamefont {S.}~\bibnamefont {Franke-Arnold}}, \ and\ \bibinfo {author} {\bibfnamefont {L.}~\bibnamefont {Chen}},\ }\href@noop {} {\bibfield  {journal} {\bibinfo  {journal} {Photonics Res.}\ }\textbf {\bibinfo {volume} {10}},\ \bibinfo {pages} {2223} (\bibinfo {year} {2022})}\BibitemShut {NoStop}%
\bibitem [{\citenamefont {Leach}\ \emph {et~al.}(2002)\citenamefont {Leach}, \citenamefont {Padgett}, \citenamefont {Barnett}, \citenamefont {Franke-Arnold},\ and\ \citenamefont {Courtial}}]{leach2002measuring}%
  \BibitemOpen
  \bibfield  {author} {\bibinfo {author} {\bibfnamefont {J.}~\bibnamefont {Leach}}, \bibinfo {author} {\bibfnamefont {M.~J.}\ \bibnamefont {Padgett}}, \bibinfo {author} {\bibfnamefont {S.~M.}\ \bibnamefont {Barnett}}, \bibinfo {author} {\bibfnamefont {S.}~\bibnamefont {Franke-Arnold}}, \ and\ \bibinfo {author} {\bibfnamefont {J.}~\bibnamefont {Courtial}},\ }\href@noop {} {\bibfield  {journal} {\bibinfo  {journal} {Phys. Rev. Lett.}\ }\textbf {\bibinfo {volume} {88}},\ \bibinfo {pages} {257901} (\bibinfo {year} {2002})}\BibitemShut {NoStop}%
\bibitem [{\citenamefont {Leach}\ \emph {et~al.}(2004)\citenamefont {Leach}, \citenamefont {Courtial}, \citenamefont {Skeldon}, \citenamefont {Barnett}, \citenamefont {Franke-Arnold},\ and\ \citenamefont {Padgett}}]{leach2004interferometric}%
  \BibitemOpen
  \bibfield  {author} {\bibinfo {author} {\bibfnamefont {J.}~\bibnamefont {Leach}}, \bibinfo {author} {\bibfnamefont {J.}~\bibnamefont {Courtial}}, \bibinfo {author} {\bibfnamefont {K.}~\bibnamefont {Skeldon}}, \bibinfo {author} {\bibfnamefont {S.~M.}\ \bibnamefont {Barnett}}, \bibinfo {author} {\bibfnamefont {S.}~\bibnamefont {Franke-Arnold}}, \ and\ \bibinfo {author} {\bibfnamefont {M.~J.}\ \bibnamefont {Padgett}},\ }\href@noop {} {\bibfield  {journal} {\bibinfo  {journal} {Phys. Rev. Lett.}\ }\textbf {\bibinfo {volume} {92}},\ \bibinfo {pages} {013601} (\bibinfo {year} {2004})}\BibitemShut {NoStop}%
\bibitem [{\citenamefont {Zhang}\ \emph {et~al.}(2014)\citenamefont {Zhang}, \citenamefont {Qi}, \citenamefont {Zhou},\ and\ \citenamefont {Chen}}]{zhang2014mimicking}%
  \BibitemOpen
  \bibfield  {author} {\bibinfo {author} {\bibfnamefont {W.}~\bibnamefont {Zhang}}, \bibinfo {author} {\bibfnamefont {Q.}~\bibnamefont {Qi}}, \bibinfo {author} {\bibfnamefont {J.}~\bibnamefont {Zhou}}, \ and\ \bibinfo {author} {\bibfnamefont {L.}~\bibnamefont {Chen}},\ }\href@noop {} {\bibfield  {journal} {\bibinfo  {journal} {Phys. Rev. Lett.}\ }\textbf {\bibinfo {volume} {112}},\ \bibinfo {pages} {153601} (\bibinfo {year} {2014})}\BibitemShut {NoStop}%
\bibitem [{\citenamefont {Berkhout}\ and\ \citenamefont {Beijersbergen}(2008)}]{berkhout2008method}%
  \BibitemOpen
  \bibfield  {author} {\bibinfo {author} {\bibfnamefont {G.~C.}\ \bibnamefont {Berkhout}}\ and\ \bibinfo {author} {\bibfnamefont {M.~W.}\ \bibnamefont {Beijersbergen}},\ }\href@noop {} {\bibfield  {journal} {\bibinfo  {journal} {Phys. Rev. Lett.}\ }\textbf {\bibinfo {volume} {101}},\ \bibinfo {pages} {100801} (\bibinfo {year} {2008})}\BibitemShut {NoStop}%
\bibitem [{\citenamefont {Hickmann}\ \emph {et~al.}(2010)\citenamefont {Hickmann}, \citenamefont {Fonseca}, \citenamefont {Soares},\ and\ \citenamefont {Ch{\'a}vez-Cerda}}]{hickmann2010unveiling}%
  \BibitemOpen
  \bibfield  {author} {\bibinfo {author} {\bibfnamefont {J.}~\bibnamefont {Hickmann}}, \bibinfo {author} {\bibfnamefont {E.}~\bibnamefont {Fonseca}}, \bibinfo {author} {\bibfnamefont {W.}~\bibnamefont {Soares}}, \ and\ \bibinfo {author} {\bibfnamefont {S.}~\bibnamefont {Ch{\'a}vez-Cerda}},\ }\href@noop {} {\bibfield  {journal} {\bibinfo  {journal} {Phys. Rev. Lett.}\ }\textbf {\bibinfo {volume} {105}},\ \bibinfo {pages} {053904} (\bibinfo {year} {2010})}\BibitemShut {NoStop}%
\bibitem [{\citenamefont {Berkhout}\ \emph {et~al.}(2010)\citenamefont {Berkhout}, \citenamefont {Lavery}, \citenamefont {Courtial}, \citenamefont {Beijersbergen},\ and\ \citenamefont {Padgett}}]{berkhout2010efficient}%
  \BibitemOpen
  \bibfield  {author} {\bibinfo {author} {\bibfnamefont {G.~C.}\ \bibnamefont {Berkhout}}, \bibinfo {author} {\bibfnamefont {M.~P.}\ \bibnamefont {Lavery}}, \bibinfo {author} {\bibfnamefont {J.}~\bibnamefont {Courtial}}, \bibinfo {author} {\bibfnamefont {M.~W.}\ \bibnamefont {Beijersbergen}}, \ and\ \bibinfo {author} {\bibfnamefont {M.~J.}\ \bibnamefont {Padgett}},\ }\href@noop {} {\bibfield  {journal} {\bibinfo  {journal} {Phys. Rev. Lett.}\ }\textbf {\bibinfo {volume} {105}},\ \bibinfo {pages} {153601} (\bibinfo {year} {2010})}\BibitemShut {NoStop}%
\bibitem [{\citenamefont {Mirhosseini}\ \emph {et~al.}(2013)\citenamefont {Mirhosseini}, \citenamefont {Malik}, \citenamefont {Shi},\ and\ \citenamefont {Boyd}}]{mirhosseini2013efficient}%
  \BibitemOpen
  \bibfield  {author} {\bibinfo {author} {\bibfnamefont {M.}~\bibnamefont {Mirhosseini}}, \bibinfo {author} {\bibfnamefont {M.}~\bibnamefont {Malik}}, \bibinfo {author} {\bibfnamefont {Z.}~\bibnamefont {Shi}}, \ and\ \bibinfo {author} {\bibfnamefont {R.~W.}\ \bibnamefont {Boyd}},\ }\href@noop {} {\bibfield  {journal} {\bibinfo  {journal} {Nat. Commun.}\ }\textbf {\bibinfo {volume} {4}},\ \bibinfo {pages} {2781} (\bibinfo {year} {2013})}\BibitemShut {NoStop}%
\bibitem [{\citenamefont {Wen}\ \emph {et~al.}(2018)\citenamefont {Wen}, \citenamefont {Chremmos}, \citenamefont {Chen}, \citenamefont {Zhu}, \citenamefont {Zhang},\ and\ \citenamefont {Yu}}]{wen2018spiral}%
  \BibitemOpen
  \bibfield  {author} {\bibinfo {author} {\bibfnamefont {Y.}~\bibnamefont {Wen}}, \bibinfo {author} {\bibfnamefont {I.}~\bibnamefont {Chremmos}}, \bibinfo {author} {\bibfnamefont {Y.}~\bibnamefont {Chen}}, \bibinfo {author} {\bibfnamefont {J.}~\bibnamefont {Zhu}}, \bibinfo {author} {\bibfnamefont {Y.}~\bibnamefont {Zhang}}, \ and\ \bibinfo {author} {\bibfnamefont {S.}~\bibnamefont {Yu}},\ }\href@noop {} {\bibfield  {journal} {\bibinfo  {journal} {Phys. Rev. Lett.}\ }\textbf {\bibinfo {volume} {120}},\ \bibinfo {pages} {193904} (\bibinfo {year} {2018})}\BibitemShut {NoStop}%
\bibitem [{\citenamefont {Liu}\ \emph {et~al.}(2019)\citenamefont {Liu}, \citenamefont {Yan}, \citenamefont {Liu},\ and\ \citenamefont {Chen}}]{liu2019superhigh}%
  \BibitemOpen
  \bibfield  {author} {\bibinfo {author} {\bibfnamefont {Z.}~\bibnamefont {Liu}}, \bibinfo {author} {\bibfnamefont {S.}~\bibnamefont {Yan}}, \bibinfo {author} {\bibfnamefont {H.}~\bibnamefont {Liu}}, \ and\ \bibinfo {author} {\bibfnamefont {X.}~\bibnamefont {Chen}},\ }\href@noop {} {\bibfield  {journal} {\bibinfo  {journal} {Phys. Rev. Lett.}\ }\textbf {\bibinfo {volume} {123}},\ \bibinfo {pages} {183902} (\bibinfo {year} {2019})}\BibitemShut {NoStop}%
\bibitem [{\citenamefont {Dudley}\ \emph {et~al.}(2013)\citenamefont {Dudley}, \citenamefont {Mhlanga}, \citenamefont {Lavery}, \citenamefont {McDonald}, \citenamefont {Roux}, \citenamefont {Padgett},\ and\ \citenamefont {Forbes}}]{dudley2013efficient}%
  \BibitemOpen
  \bibfield  {author} {\bibinfo {author} {\bibfnamefont {A.}~\bibnamefont {Dudley}}, \bibinfo {author} {\bibfnamefont {T.}~\bibnamefont {Mhlanga}}, \bibinfo {author} {\bibfnamefont {M.}~\bibnamefont {Lavery}}, \bibinfo {author} {\bibfnamefont {A.}~\bibnamefont {McDonald}}, \bibinfo {author} {\bibfnamefont {F.~S.}\ \bibnamefont {Roux}}, \bibinfo {author} {\bibfnamefont {M.}~\bibnamefont {Padgett}}, \ and\ \bibinfo {author} {\bibfnamefont {A.}~\bibnamefont {Forbes}},\ }\href@noop {} {\bibfield  {journal} {\bibinfo  {journal} {Opt. Express}\ }\textbf {\bibinfo {volume} {21}},\ \bibinfo {pages} {165} (\bibinfo {year} {2013})}\BibitemShut {NoStop}%
\bibitem [{\citenamefont {Trichili}\ \emph {et~al.}(2014)\citenamefont {Trichili}, \citenamefont {Mhlanga}, \citenamefont {Ismail}, \citenamefont {Roux}, \citenamefont {McLaren}, \citenamefont {Zghal},\ and\ \citenamefont {Forbes}}]{trichili2014detection}%
  \BibitemOpen
  \bibfield  {author} {\bibinfo {author} {\bibfnamefont {A.}~\bibnamefont {Trichili}}, \bibinfo {author} {\bibfnamefont {T.}~\bibnamefont {Mhlanga}}, \bibinfo {author} {\bibfnamefont {Y.}~\bibnamefont {Ismail}}, \bibinfo {author} {\bibfnamefont {F.~S.}\ \bibnamefont {Roux}}, \bibinfo {author} {\bibfnamefont {M.}~\bibnamefont {McLaren}}, \bibinfo {author} {\bibfnamefont {M.}~\bibnamefont {Zghal}}, \ and\ \bibinfo {author} {\bibfnamefont {A.}~\bibnamefont {Forbes}},\ }\href@noop {} {\bibfield  {journal} {\bibinfo  {journal} {Opt. Express}\ }\textbf {\bibinfo {volume} {22}},\ \bibinfo {pages} {17553} (\bibinfo {year} {2014})}\BibitemShut {NoStop}%
\bibitem [{\citenamefont {Fickler}\ \emph {et~al.}(2017)\citenamefont {Fickler}, \citenamefont {Ginoya},\ and\ \citenamefont {Boyd}}]{fickler2017custom}%
  \BibitemOpen
  \bibfield  {author} {\bibinfo {author} {\bibfnamefont {R.}~\bibnamefont {Fickler}}, \bibinfo {author} {\bibfnamefont {M.}~\bibnamefont {Ginoya}}, \ and\ \bibinfo {author} {\bibfnamefont {R.~W.}\ \bibnamefont {Boyd}},\ }\href@noop {} {\bibfield  {journal} {\bibinfo  {journal} {Phys. Rev. B}\ }\textbf {\bibinfo {volume} {95}},\ \bibinfo {pages} {161108} (\bibinfo {year} {2017})}\BibitemShut {NoStop}%
\bibitem [{\citenamefont {Fickler}\ \emph {et~al.}(2020)\citenamefont {Fickler}, \citenamefont {Bouchard}, \citenamefont {Giese}, \citenamefont {Grillo}, \citenamefont {Leuchs},\ and\ \citenamefont {Karimi}}]{fickler2020full}%
  \BibitemOpen
  \bibfield  {author} {\bibinfo {author} {\bibfnamefont {R.}~\bibnamefont {Fickler}}, \bibinfo {author} {\bibfnamefont {F.}~\bibnamefont {Bouchard}}, \bibinfo {author} {\bibfnamefont {E.}~\bibnamefont {Giese}}, \bibinfo {author} {\bibfnamefont {V.}~\bibnamefont {Grillo}}, \bibinfo {author} {\bibfnamefont {G.}~\bibnamefont {Leuchs}}, \ and\ \bibinfo {author} {\bibfnamefont {E.}~\bibnamefont {Karimi}},\ }\href@noop {} {\bibfield  {journal} {\bibinfo  {journal} {J. Opt.}\ }\textbf {\bibinfo {volume} {22}},\ \bibinfo {pages} {024001} (\bibinfo {year} {2020})}\BibitemShut {NoStop}%
\bibitem [{\citenamefont {Gu}\ \emph {et~al.}(2018)\citenamefont {Gu}, \citenamefont {Krenn}, \citenamefont {Erhard},\ and\ \citenamefont {Zeilinger}}]{gu2018gouy}%
  \BibitemOpen
  \bibfield  {author} {\bibinfo {author} {\bibfnamefont {X.}~\bibnamefont {Gu}}, \bibinfo {author} {\bibfnamefont {M.}~\bibnamefont {Krenn}}, \bibinfo {author} {\bibfnamefont {M.}~\bibnamefont {Erhard}}, \ and\ \bibinfo {author} {\bibfnamefont {A.}~\bibnamefont {Zeilinger}},\ }\href@noop {} {\bibfield  {journal} {\bibinfo  {journal} {Phys. Rev. Lett.}\ }\textbf {\bibinfo {volume} {120}},\ \bibinfo {pages} {103601} (\bibinfo {year} {2018})}\BibitemShut {NoStop}%
\bibitem [{\citenamefont {Zhou}\ \emph {et~al.}(2017)\citenamefont {Zhou}, \citenamefont {Mirhosseini}, \citenamefont {Fu}, \citenamefont {Zhao}, \citenamefont {Rafsanjani}, \citenamefont {Willner},\ and\ \citenamefont {Boyd}}]{zhou2017sorting}%
  \BibitemOpen
  \bibfield  {author} {\bibinfo {author} {\bibfnamefont {Y.}~\bibnamefont {Zhou}}, \bibinfo {author} {\bibfnamefont {M.}~\bibnamefont {Mirhosseini}}, \bibinfo {author} {\bibfnamefont {D.}~\bibnamefont {Fu}}, \bibinfo {author} {\bibfnamefont {J.}~\bibnamefont {Zhao}}, \bibinfo {author} {\bibfnamefont {S.~M.~H.}\ \bibnamefont {Rafsanjani}}, \bibinfo {author} {\bibfnamefont {A.~E.}\ \bibnamefont {Willner}}, \ and\ \bibinfo {author} {\bibfnamefont {R.~W.}\ \bibnamefont {Boyd}},\ }\href@noop {} {\bibfield  {journal} {\bibinfo  {journal} {Phys. Rev. Lett.}\ }\textbf {\bibinfo {volume} {119}},\ \bibinfo {pages} {263602} (\bibinfo {year} {2017})}\BibitemShut {NoStop}%
\bibitem [{\citenamefont {Fu}\ \emph {et~al.}(2018)\citenamefont {Fu}, \citenamefont {Zhou}, \citenamefont {Qi}, \citenamefont {Oliver}, \citenamefont {Wang}, \citenamefont {Rafsanjani}, \citenamefont {Zhao}, \citenamefont {Mirhosseini}, \citenamefont {Shi}, \citenamefont {Zhang} \emph {et~al.}}]{fu2018realization}%
  \BibitemOpen
  \bibfield  {author} {\bibinfo {author} {\bibfnamefont {D.}~\bibnamefont {Fu}}, \bibinfo {author} {\bibfnamefont {Y.}~\bibnamefont {Zhou}}, \bibinfo {author} {\bibfnamefont {R.}~\bibnamefont {Qi}}, \bibinfo {author} {\bibfnamefont {S.}~\bibnamefont {Oliver}}, \bibinfo {author} {\bibfnamefont {Y.}~\bibnamefont {Wang}}, \bibinfo {author} {\bibfnamefont {S.~M.~H.}\ \bibnamefont {Rafsanjani}}, \bibinfo {author} {\bibfnamefont {J.}~\bibnamefont {Zhao}}, \bibinfo {author} {\bibfnamefont {M.}~\bibnamefont {Mirhosseini}}, \bibinfo {author} {\bibfnamefont {Z.}~\bibnamefont {Shi}}, \bibinfo {author} {\bibfnamefont {P.}~\bibnamefont {Zhang}},  \emph {et~al.},\ }\href@noop {} {\bibfield  {journal} {\bibinfo  {journal} {Opt. Express}\ }\textbf {\bibinfo {volume} {26}},\ \bibinfo {pages} {33057} (\bibinfo {year} {2018})}\BibitemShut {NoStop}%
\bibitem [{\citenamefont {Fontaine}\ \emph {et~al.}(2019)\citenamefont {Fontaine}, \citenamefont {Ryf}, \citenamefont {Chen}, \citenamefont {Neilson}, \citenamefont {Kim},\ and\ \citenamefont {Carpenter}}]{fontaine2019laguerre}%
  \BibitemOpen
  \bibfield  {author} {\bibinfo {author} {\bibfnamefont {N.~K.}\ \bibnamefont {Fontaine}}, \bibinfo {author} {\bibfnamefont {R.}~\bibnamefont {Ryf}}, \bibinfo {author} {\bibfnamefont {H.}~\bibnamefont {Chen}}, \bibinfo {author} {\bibfnamefont {D.~T.}\ \bibnamefont {Neilson}}, \bibinfo {author} {\bibfnamefont {K.}~\bibnamefont {Kim}}, \ and\ \bibinfo {author} {\bibfnamefont {J.}~\bibnamefont {Carpenter}},\ }\href@noop {} {\bibfield  {journal} {\bibinfo  {journal} {Nat. Commun.}\ }\textbf {\bibinfo {volume} {10}},\ \bibinfo {pages} {1865} (\bibinfo {year} {2019})}\BibitemShut {NoStop}%
\bibitem [{\citenamefont {Plick}\ and\ \citenamefont {Krenn}(2015)}]{plick2015physical}%
  \BibitemOpen
  \bibfield  {author} {\bibinfo {author} {\bibfnamefont {W.~N.}\ \bibnamefont {Plick}}\ and\ \bibinfo {author} {\bibfnamefont {M.}~\bibnamefont {Krenn}},\ }\href@noop {} {\bibfield  {journal} {\bibinfo  {journal} {Phys. Rev. A}\ }\textbf {\bibinfo {volume} {92}},\ \bibinfo {pages} {063841} (\bibinfo {year} {2015})}\BibitemShut {NoStop}%
\bibitem [{\citenamefont {Sahu}\ \emph {et~al.}(2018)\citenamefont {Sahu}, \citenamefont {Chaudhary}, \citenamefont {Khare}, \citenamefont {Bhattacharya}, \citenamefont {Wanare},\ and\ \citenamefont {Jha}}]{sahu2018angular}%
  \BibitemOpen
  \bibfield  {author} {\bibinfo {author} {\bibfnamefont {R.}~\bibnamefont {Sahu}}, \bibinfo {author} {\bibfnamefont {S.}~\bibnamefont {Chaudhary}}, \bibinfo {author} {\bibfnamefont {K.}~\bibnamefont {Khare}}, \bibinfo {author} {\bibfnamefont {M.}~\bibnamefont {Bhattacharya}}, \bibinfo {author} {\bibfnamefont {H.}~\bibnamefont {Wanare}}, \ and\ \bibinfo {author} {\bibfnamefont {A.~K.}\ \bibnamefont {Jha}},\ }\href@noop {} {\bibfield  {journal} {\bibinfo  {journal} {Opt. Express}\ }\textbf {\bibinfo {volume} {26}},\ \bibinfo {pages} {8709} (\bibinfo {year} {2018})}\BibitemShut {NoStop}%
\bibitem [{\citenamefont {Zhou}\ \emph {et~al.}(2022)\citenamefont {Zhou}, \citenamefont {Pu},\ and\ \citenamefont {Wang}}]{zhou2022orbital}%
  \BibitemOpen
  \bibfield  {author} {\bibinfo {author} {\bibfnamefont {J.}~\bibnamefont {Zhou}}, \bibinfo {author} {\bibfnamefont {H.}~\bibnamefont {Pu}}, \ and\ \bibinfo {author} {\bibfnamefont {Q.}~\bibnamefont {Wang}},\ }\href@noop {} {\bibfield  {journal} {\bibinfo  {journal} {Opt. Express}\ }\textbf {\bibinfo {volume} {30}},\ \bibinfo {pages} {9703} (\bibinfo {year} {2022})}\BibitemShut {NoStop}%
\bibitem [{\citenamefont {Lv}\ \emph {et~al.}(2022)\citenamefont {Lv}, \citenamefont {Shang}, \citenamefont {Fu}, \citenamefont {Huang}, \citenamefont {Gao},\ and\ \citenamefont {Gao}}]{lv2022sorting}%
  \BibitemOpen
  \bibfield  {author} {\bibinfo {author} {\bibfnamefont {Y.}~\bibnamefont {Lv}}, \bibinfo {author} {\bibfnamefont {Z.}~\bibnamefont {Shang}}, \bibinfo {author} {\bibfnamefont {S.}~\bibnamefont {Fu}}, \bibinfo {author} {\bibfnamefont {L.}~\bibnamefont {Huang}}, \bibinfo {author} {\bibfnamefont {L.}~\bibnamefont {Gao}}, \ and\ \bibinfo {author} {\bibfnamefont {C.}~\bibnamefont {Gao}},\ }\href@noop {} {\bibfield  {journal} {\bibinfo  {journal} {Opt. Lett.}\ }\textbf {\bibinfo {volume} {47}},\ \bibinfo {pages} {5032} (\bibinfo {year} {2022})}\BibitemShut {NoStop}%
\bibitem [{\citenamefont {Guo}\ \emph {et~al.}(2021)\citenamefont {Guo}, \citenamefont {Zhang}, \citenamefont {Pu}, \citenamefont {He}, \citenamefont {Jin}, \citenamefont {Xu}, \citenamefont {Zhang}, \citenamefont {Gao},\ and\ \citenamefont {Luo}}]{guo2021spin}%
  \BibitemOpen
  \bibfield  {author} {\bibinfo {author} {\bibfnamefont {Y.}~\bibnamefont {Guo}}, \bibinfo {author} {\bibfnamefont {S.}~\bibnamefont {Zhang}}, \bibinfo {author} {\bibfnamefont {M.}~\bibnamefont {Pu}}, \bibinfo {author} {\bibfnamefont {Q.}~\bibnamefont {He}}, \bibinfo {author} {\bibfnamefont {J.}~\bibnamefont {Jin}}, \bibinfo {author} {\bibfnamefont {M.}~\bibnamefont {Xu}}, \bibinfo {author} {\bibfnamefont {Y.}~\bibnamefont {Zhang}}, \bibinfo {author} {\bibfnamefont {P.}~\bibnamefont {Gao}}, \ and\ \bibinfo {author} {\bibfnamefont {X.}~\bibnamefont {Luo}},\ }\href@noop {} {\bibfield  {journal} {\bibinfo  {journal} {Light: Sci. Appl.}\ }\textbf {\bibinfo {volume} {10}},\ \bibinfo {pages} {63} (\bibinfo {year} {2021})}\BibitemShut {NoStop}%
\bibitem [{\citenamefont {Wang}\ \emph {et~al.}(2024)\citenamefont {Wang}, \citenamefont {Wu}, \citenamefont {Lin}, \citenamefont {Zhong}, \citenamefont {Zhu},\ and\ \citenamefont {Chen}}]{wangmetasurface}%
  \BibitemOpen
  \bibfield  {author} {\bibinfo {author} {\bibfnamefont {B.}~\bibnamefont {Wang}}, \bibinfo {author} {\bibfnamefont {L.}~\bibnamefont {Wu}}, \bibinfo {author} {\bibfnamefont {Z.}~\bibnamefont {Lin}}, \bibinfo {author} {\bibfnamefont {W.}~\bibnamefont {Zhong}}, \bibinfo {author} {\bibfnamefont {Z.}~\bibnamefont {Zhu}}, \ and\ \bibinfo {author} {\bibfnamefont {Y.}~\bibnamefont {Chen}},\ }\href@noop {} {\bibfield  {journal} {\bibinfo  {journal} {Laser Photonics Rev.}\ ,\ \bibinfo {pages} {2400113}} (\bibinfo {year} {2024})}\BibitemShut {NoStop}%
\bibitem [{\citenamefont {Ahmed}\ \emph {et~al.}(2022)\citenamefont {Ahmed}, \citenamefont {Kim}, \citenamefont {Zhang}, \citenamefont {Intaravanne}, \citenamefont {Jang}, \citenamefont {Rho}, \citenamefont {Chen},\ and\ \citenamefont {Chen}}]{ahmed2022optical}%
  \BibitemOpen
  \bibfield  {author} {\bibinfo {author} {\bibfnamefont {H.}~\bibnamefont {Ahmed}}, \bibinfo {author} {\bibfnamefont {H.}~\bibnamefont {Kim}}, \bibinfo {author} {\bibfnamefont {Y.}~\bibnamefont {Zhang}}, \bibinfo {author} {\bibfnamefont {Y.}~\bibnamefont {Intaravanne}}, \bibinfo {author} {\bibfnamefont {J.}~\bibnamefont {Jang}}, \bibinfo {author} {\bibfnamefont {J.}~\bibnamefont {Rho}}, \bibinfo {author} {\bibfnamefont {S.}~\bibnamefont {Chen}}, \ and\ \bibinfo {author} {\bibfnamefont {X.}~\bibnamefont {Chen}},\ }\href@noop {} {\bibfield  {journal} {\bibinfo  {journal} {Nanophotonics}\ }\textbf {\bibinfo {volume} {11}},\ \bibinfo {pages} {941} (\bibinfo {year} {2022})}\BibitemShut {NoStop}%
\bibitem [{\citenamefont {Cheng}\ \emph {et~al.}(2022)\citenamefont {Cheng}, \citenamefont {Sha}, \citenamefont {Zhang}, \citenamefont {Chen}, \citenamefont {Qu}, \citenamefont {Song}, \citenamefont {Yu},\ and\ \citenamefont {Xiao}}]{cheng2022ultracompact}%
  \BibitemOpen
  \bibfield  {author} {\bibinfo {author} {\bibfnamefont {J.}~\bibnamefont {Cheng}}, \bibinfo {author} {\bibfnamefont {X.}~\bibnamefont {Sha}}, \bibinfo {author} {\bibfnamefont {H.}~\bibnamefont {Zhang}}, \bibinfo {author} {\bibfnamefont {Q.}~\bibnamefont {Chen}}, \bibinfo {author} {\bibfnamefont {G.}~\bibnamefont {Qu}}, \bibinfo {author} {\bibfnamefont {Q.}~\bibnamefont {Song}}, \bibinfo {author} {\bibfnamefont {S.}~\bibnamefont {Yu}}, \ and\ \bibinfo {author} {\bibfnamefont {S.}~\bibnamefont {Xiao}},\ }\href@noop {} {\bibfield  {journal} {\bibinfo  {journal} {Nano Lett.}\ }\textbf {\bibinfo {volume} {22}},\ \bibinfo {pages} {3993} (\bibinfo {year} {2022})}\BibitemShut {NoStop}%
\end{thebibliography}
%

\end{document}